\documentclass[
aps,prapplied,showpacs,superscriptaddress,numerical,amsmath,amssymb,floatfix,reprint,longbibliography]{revtex4-2}
\usepackage[T1]{fontenc}
\usepackage[utf8]{inputenc}
\usepackage[
pdffitwindow=true,
colorlinks=true,
frenchlinks=false,
linkcolor=blue,
anchorcolor=blue,
citecolor=blue,
filecolor=blue,
urlcolor=blue,
bookmarks=true,
bookmarksopen=true,
bookmarksnumbered=true,
bookmarksopenlevel=1,
plainpages=false,
pdfpagelayout=TwoPageLeft,
pdfpagelabels=true,
breaklinks]{hyperref}
\usepackage[main=english]{babel}
\usepackage[dvipsnames]{xcolor}
\usepackage[per-mode=symbol,separate-uncertainty]{siunitx}
\usepackage{graphicx}
\usepackage{dcolumn}
\usepackage{comment}
\usepackage{color}
\usepackage{graphicx,wrapfig,lipsum}
\usepackage{bm}
\usepackage{physics}
\usepackage{color}
\usepackage{ulem}
\usepackage{float}
\usepackage{chemformula}
\usepackage{todonotes}
\usepackage{cleveref}

\usepackage{soul} % for strike out text
\DeclareSIUnit{\nT}{\nano\tesla}
\DeclareSIUnit{\uT}{\micro\tesla}
\DeclareSIUnit{\mT}{\milli\tesla}
\DeclareSIUnit{\T}{\tesla}
\DeclareSIUnit{\mK}{\milli\kelvin}

\newcommand{\aref}[1]{\autoref{#1}}

\begin{document}

%\preprint{AAPM/123-QED}

\title{Experimental setup for the combined study of spin ensembles and superconducting quantum circuits }
%\thanks{Footnote to title of article.}
%NOTE Generation of weak magnetic fields in the vicinity of superconducting flux-tunable qubits

\author{Lukas~Vogl}
\affiliation{Walther-Mei{\ss}ner-Institut, Bayerische Akademie der Wissenschaften, 85748 Garching, Germany} 
\affiliation{TUM School of Natural Sciences, Physics Department,Technical University of Munich, 85748 Garching, Germany}
\affiliation{Munich Center for Quantum Science and Technology (MCQST), 80799 Munich, Germany}
% \email{lukas.vogl@tum.de}
\author{Gerhard~B.~P.~Huber}
\affiliation{Walther-Mei{\ss}ner-Institut, Bayerische Akademie der Wissenschaften, 85748 Garching, Germany} 
\affiliation{TUM School of Natural Sciences, Physics Department,Technical University of Munich, 85748 Garching, Germany}
\affiliation{Munich Center for Quantum Science and Technology (MCQST), 80799 Munich, Germany}
\author{Ana~Strini\'{c}}%
\affiliation{Walther-Mei{\ss}ner-Institut, Bayerische Akademie der Wissenschaften, 85748 Garching, Germany} 
\affiliation{TUM School of Natural Sciences, Physics Department,Technical University of Munich, 85748 Garching, Germany}
\affiliation{Munich Center for Quantum Science and Technology (MCQST), 80799 Munich, Germany}
\author{Achim~Marx}
\affiliation{Walther-Mei{\ss}ner-Institut, Bayerische Akademie der Wissenschaften, 85748 Garching, Germany} 
\affiliation{TUM School of Natural Sciences, Physics Department,Technical University of Munich, 85748 Garching, Germany}
\affiliation{Munich Center for Quantum Science and Technology (MCQST), 80799 Munich, Germany}
\author{Stefan~Filipp}
\affiliation{Walther-Mei{\ss}ner-Institut, Bayerische Akademie der Wissenschaften, 85748 Garching, Germany} 
\affiliation{TUM School of Natural Sciences, Physics Department,Technical University of Munich, 85748 Garching, Germany}
\affiliation{Munich Center for Quantum Science and Technology (MCQST), 80799 Munich, Germany}
\author{Kirill~G.~Fedorov}
\affiliation{Walther-Mei{\ss}ner-Institut, Bayerische Akademie der Wissenschaften, 85748 Garching, Germany} 
\affiliation{TUM School of Natural Sciences, Physics Department,Technical University of Munich, 85748 Garching, Germany}
\affiliation{Munich Center for Quantum Science and Technology (MCQST), 80799 Munich, Germany}
\author{Rudolf~Gross}
\affiliation{Walther-Mei{\ss}ner-Institut, Bayerische Akademie der Wissenschaften, 85748 Garching, Germany} 
\affiliation{TUM School of Natural Sciences, Physics Department,Technical University of Munich, 85748 Garching, Germany}
\affiliation{Munich Center for Quantum Science and Technology (MCQST), 80799 Munich, Germany}
\author{Nadezhda~P.~Kukharchyk}
 \email{Nadezhda.Kukharchyk@wmi.badw.de}
 %\homepage{http://www.Second.institution.edu/~Charlie.Author.}
\affiliation{Walther-Mei{\ss}ner-Institut, Bayerische Akademie der Wissenschaften, 85748 Garching, Germany} 
\affiliation{TUM School of Natural Sciences, Physics Department,Technical University of Munich, 85748 Garching, Germany}
\affiliation{Munich Center for Quantum Science and Technology (MCQST), 80799 Munich, Germany}
\date{\today}% It is always \today, today,
 
\begin{abstract}
 
% Alternative abstract: 
A hybrid quantum computing architecture combining quantum processors and quantum memory units allows for exploiting each component's unique properties to enhance the overall performance of the total system. However, superconducting qubits are highly sensitive to magnetic fields, while spin ensembles require finite fields for control, creating a major integration challenge. In this work, we demonstrate the first experimental setup that satisfies these constraints and provides verified qubit stability. Our cryogenic setup comprises two spatially and magnetically decoupled sample volumes inside a single dilution refrigerator: one hosting flux-tunable superconducting qubits and the other a spin ensemble equipped with a superconducting solenoid generating fields up to \SI{50}{\milli\tesla}. 
We show that several layers of Cryophy\textsuperscript{\textregistered} shielding and an additional superconducting aluminum shield suppress magnetic crosstalk by more than eight orders of magnitude, ensuring stability of the qubit's performance. 
Moreover, the operation of the solenoid adds minimal thermal load on the relevant stages of the dilution refrigerator. 
Our results enable scalable hybrid quantum architectures with low-loss integration, marking a key step toward scalable hybrid quantum computing platforms.
\end{abstract}

\maketitle
%\tableofcontents
%\begin{quotation}
%Mostly place holder texts and images
%\end{quotation}

% - Comparison to other works\\
% - Discuss the issues related to SC shields. The effect of the sc shield on the modeling; flux-trapping\\
% - Numerical estimates of the thermal load: reduce or remove.\\

\label{sec:intro}
\section{Introduction}
The development of hybrid quantum devices that integrate superconducting quantum circuits with quantum systems based on electron spins is driven by the rapid upscaling in superconducting quantum computing and the inherent problems of a limited on-chip qubit density and missing quantum memory~\cite{Kurizki2015,Gouzien2021}. A possible solution for the limited qubit density is the development of quantum bus systems between chiplets within the same refrigerator or even between processors operated in spatially separated units~\cite{Magnard2020cryolink, Renger2025cryolink}, allowing for distributed quantum computing. For the integration of superconducting quantum processors with spin ensemble-based quantum memories within the same dilution refrigerator, two problems have to be solved. First, one has to develop efficient coupling schemes between independently controlled spin ensembles and superconducting quantum circuits. To this end, the realization of a low-loss interface bus between the two subsystems is mandatory. Second, as discussed here, one has to realize a cryogenic setup, allowing the stable operation of both subsystems, having incompatible operation conditions regarding the magnetic field environment.

Solid-state spin-ensemble quantum memories~\cite{Tyryshkin.2011,Bottger.2006,Kukharchyk.2018} typically operate under moderate to high magnetic fields, namely \SI{300}{\milli\tesla} -- \SI{1.5}{\tesla}. At the same time, such fields perturb the operation of superconducting quantum circuits~\cite{Bothner.2012,Schneider.2019}. This incompatibility requires positioning both units in separate sample volumes and taking care of proper magnetic decoupling. A simple solution for keeping the perturbing fields below a tolerable threshold value is to increase the distance between the two sample volumes, as the field strength rapidly decreases with distance. Recent advances in cryogenic engineering~\cite{Magnard2020cryolink, Renger2025cryolink} offer a promising approach to interconnect two sample volumes operated at millikelvin temperatures via a low-loss cryogenic microwave link over a distance of several meters. Nevertheless, in many situations, a more compact architecture with both sample volumes integrated into the same cryogenic environment is desirable to minimize losses and efforts in cryogenic engineering~\cite{Gouzien2021}. However, as discussed here, for realizing such a hybrid architecture, one has to develop efficient magnetic shielding. 

The characteristic parameters of superconducting qubits, such as their transition frequency and coherence time, are highly susceptible to variations in the ambient magnetic field. On the one hand, this allows simple tuning of the qubit parameters of flux-tunable qubits, while on the other hand, one has to carefully remove unwanted magnetic field variations. The latter can have many causes and cover a wide frequency range. Prominent examples include two-level systems (TLS) associated with defect structures at surfaces and interfaces, the electromagnetic mode environment, and the typical electromagnetic field noise present in a lab environment. Magnetic field variations are particularly affecting qubits, which can be tuned by an applied magnetic flux. They lead to decoherence~\cite{kumar2016, Goetz_2016, wang2015, Goetz_2017, Gozzelino_2022} and the need for frequent recalibration due to variations in the magnetic environment~\cite{Schneider2019}. Therefore, clarifying the various sources of magnetic field noise and developing strategies for their mitigation is a key task in the development of scalable quantum computing. Besides the reduction of the TLS density in qubit structures, the efficient suppression of perturbing magnetic field sources is crucial for maintaining the stable operation of superconducting quantum processors. This also applies to the suppression of the magnetic fields required for the operation of a nearby spin ensemble-based quantum memory. 

The suppression of ambient magnetic fields can be achieved through either passive or active shielding methods, with the optimal approach depending on the field’s origin and the specific application. Static fields -- such as the Earth's magnetic field -- are most effectively mitigated using passive magnetic shielding composed of materials that expel or redirect magnetic flux~\cite{Wadey1956, Sumner_1987, Santos2012, Whelan_2018}.
In contrast, active magnetic field suppression is more suitable for confined sample volumes and can be integrated with systems that generate tunable magnetic fields~\cite{Brake1991, Brys2005, Afach2014, Holmes2022, Abel2023}.

Rare-earth spin ensembles are known for their long coherence times, particularly when operated at Zero First-Order Zeeman (ZEFOZ) points. These points, analogous to clock transitions, offer enhanced coherence by minimizing the system’s sensitivity to external magnetic field fluctuations. However, achieving the ZEFOZ condition requires applying a finite magnetic field and precisely controlling it, often involving fine-tuning of both its magnitude and orientation~\cite{mcauslan2012}. 

A hybrid system integrating superconducting quantum circuits with magnetically controlled spin ensembles in the same dilution refrigerator requires the realization of a large, magnetically compensated sample volume for housing the field-sensitive superconducting circuits. This requirement cannot be met solely through the use of compensation coils. It rather requires the design of a specialized magnetic shielding structure capable of confining the magnetic fields used for controlling the spin ensemble within the sample volume housing the spin system. Moreover, the magnetic field must meet strict requirements regarding homogeneity, stability, and strength to ensure proper spin ensemble operation. Finally, the nearby sample volume housing the superconducting quantum circuit must remain entirely unaffected by the magnetic field variations within the sample volume of the spin ensemble. This requires a careful magnetic decoupling between the two subsystems.
Most works concerning the qubit coherence discuss the possible shielding from static fields, such as the Earth field~\cite{Malevannaya.2025}.

In this work, we present a hybrid architecture integrating a sample volume containing a superconducting solenoid for controlling a spin ensemble with a second sample volume where superconducting circuits can be operated at a sufficiently low background field within one and the same dry dilution refrigerator~\cite{Uhlig2008}. The setup fulfills key requirements for magnetic field generation regarding stability, tunability, and homogeneity, while simultaneously providing efficient magnetic shielding to allow for the nearby operation of magnetic field-sensitive superconducting circuits. In particular, our setup enables precise magnetic field control at the position of the spin ensemble, facilitating fine-tuning to Zero First-Order Zeeman (ZEFOZ) transitions in spin ensembles~\cite{Ortu2018}. Moreover, it ensures that flux-tunable superconducting qubits remain unaffected by the applied magnetic fields, thereby establishing a viable pathway for coupling superconducting qubits to spin ensembles within a shared cryogenic environment.

\section{Hybrid Cryogenic Setup}
   
\begin{figure}[b]
    \includegraphics[width=1.0\columnwidth]{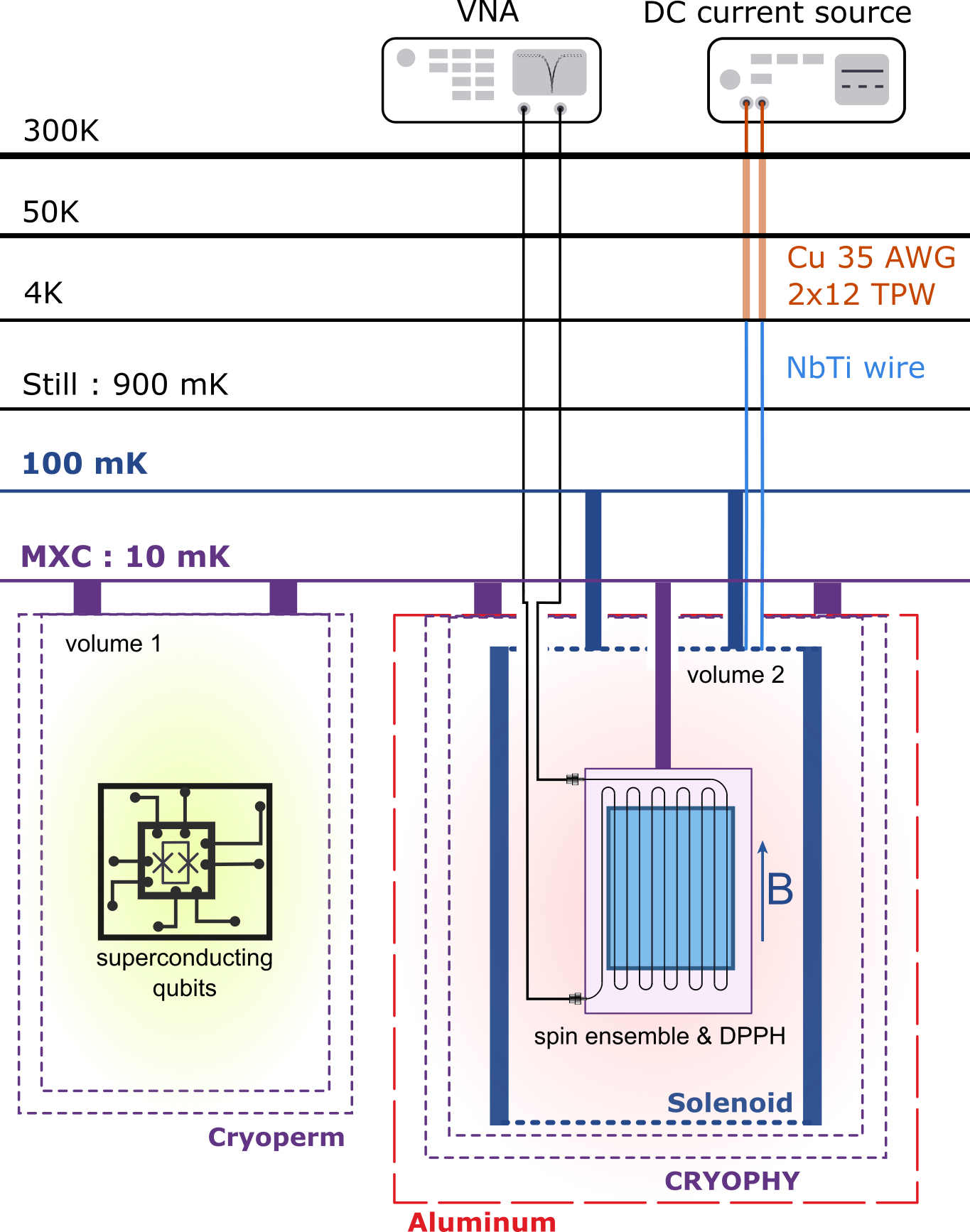} 
    \caption{\label{fig:1setup}
    Schematics of the dilution refrigerator and hybrid cryogenic setup. The refrigerator is represented by six temperature stages with additional microwave attenuation of input lines and low-noise preamplification of the output lines (not shown here for simplicity). All the magnetic shields as well as the experimental samples are thermally anchored to the mixing chamber plate (MXC) at the base temperature of less than \SI{10}{\milli\kelvin}. The superconducting solenoid magnet is thermally anchored to the intermediate cold plate at the temperature of $\simeq$\SI{100}{\milli\kelvin}. The magnet is powered via superconducting wires between the MXC- and the \SI{4}{\kelvin}-stage and via copper DC-lines between the \SI{4}{\kelvin}-stage and room temperature.}    
\end{figure}

The hybrid cryogenic setup consists of two spatially separated and independently controlled experimental cryogenic sample volumes, denoted as sample volumes~1 and~2 in the following. In the context of this experimental work, the two volumes are neither directly connected by microwave nor by low-frequency signal lines and are controlled independently of each other. The two volumes are separated by approximately \SI{250}{\milli\meter} center-to-center and are placed within a commercial Bluefors XLD~1000sl dilution refrigerator, which imposes limitations on the maximum dimensions of the two sample volumes and their spatial separation. The schematic of the assembly is shown in \Cref{fig:1setup}. 

\subsection{Sample Volume for Superconducting Quantum Circuits}

The superconducting qubits are mounted in the sample volume~1 as shown in \Cref{fig:1setup} and are protected against perturbing magnetic fields by two layers of Cryoperm$^\circledR$, which are thermally anchored to the MXC base plate. The details on the mounting of the superconducting qubits in sample volume~1 can be found in the Appendix of Ref.~\cite{Huber2025}. In this study, the transition frequency and coherence properties of a flux-tunable qubit (FTQ) are measured to assess potential influences from the magnetic fields generated in sample volume~2 housing the spin ensemble. The FTQ consists of a SQUID-loop with dimensions of \SI{13}{\micro \meter} by \SI{23}{\micro \meter}, characterized by Josephson coupling energies $E_\text{J1}/h = \SI{16.9}{\giga\hertz}$, $E_\text{J2}/h = \SI{8.4}{\giga\hertz}$ and the charging energy $E_{\text{C}}/h = \SI{160.6}{\mega\hertz}$~\cite{Huber2025}. It is placed in the lower part of the sample volume~1 with its plane aligned parallel to the axis of the solenoid.

\subsection{Sample Volume for Spin Ensemble}

The space hosting the spin ensemble is referred to as sample volume~2. It comprises a superconducting solenoid enclosed within three layers of magnetic shielding. The solenoid is thermally anchored to the \SI{100}{\milli\kelvin} stage of the dilution refrigerator, while the magnetic shields and the sample holder are connected to the MXC base plate reaching a temperature below \SI{10}{\milli\kelvin}, as depicted in the expanded three-dimensional model in \Cref{fig:2_surface_plot}(a). A detailed characterization of the spin ensembles is provided in Ref.~\cite{Strinic2025}. In this work, we focus on the design and implementation of the experimental setup.

\section{Superconducting Solenoid}
\label{subsection:solenoid_construction}

\begin{figure}[tb]
	\centering
	\includegraphics[width=1.0\columnwidth]{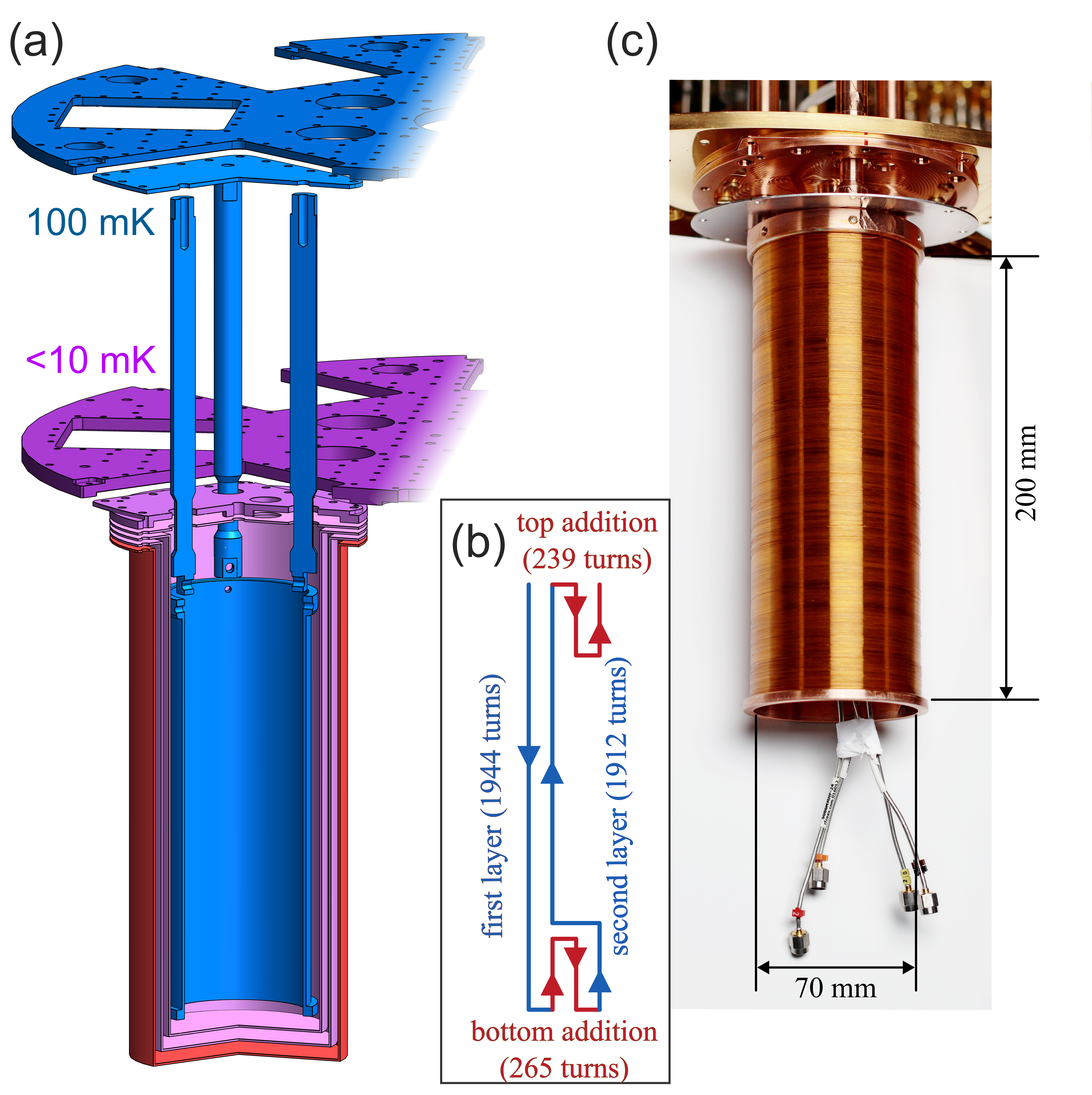}
	\caption{(a)~Three-dimensional graphic representation of the sample volume~2 and its mounting within the cryostat. The representation has an in-cut to show all inner layers of the assembly. Different colors are used for parts anchored at different temperature stages (blue: \SI{100}{\milli\kelvin} plate, purple: MXC base plate) of the refrigerator. The superconducting aluminum shield (red color) is also thermally anchored to the MXC base plate.
    (b)~Diagram of the winding pattern of the superconducting solenoid, which features the winding direction, layering and number of turns.
	(c)~Photo of the superconducting solenoid without the shielding layers when installed into the cryostat.}
	\label{fig:3_3d_overview}
\end{figure}

The magnetic field in sample volume~2 is generated by a superconducting solenoid. It is obtained by winding a superconducting NbTi wire onto a cylinder-shaped solenoid body made of copper. It has a winding length of \SI{200}{\milli\meter}, an inner diameter of \SI{70}{\milli\meter}, and wall thickness of \SI{1.5}{\milli\meter}, resulting in an outer coil diameter of \SI{73}{\milli\meter}. It is held by three copper rods, which are attached to a side-loading RF port at the \SI{100}{\milli\kelvin} thermal plate of the Bluefors~XLD~1000sl system [cf. \Cref{fig:3_3d_overview}\,(a)]. The magnetic shields are fixed at a copper flange, which is thermally anchored at the base plate of the refrigerator. All copper parts are fabricated from oxygen-free high thermal conductance copper (OFHC) with a purity of $>99.99\%$. 

The magnetic field distribution generated by a solenoid enclosed in magnetic shielding layers is strongly influenced by the type and geometry of the shielding~\cite{Morecroft1925}. To compensate for the effects of the cryogenic magnetic shielding (CMS) and to enhance the overall homogeneity of the magnetic field at the sample position, additional partial winding layers of superconducting wire were added at both the top and bottom ends of the solenoid, as illustrated in \Cref{fig:3_3d_overview}(b). This configuration results in two full and four partial winding layers. The optimal lengths of the partial layers were determined through numerical simulations exploring various length configurations. The solenoid was wound using a \SI{101}{\micro\meter}-thick superconducting NbTi wire, with the bare NbTi wire core having a diameter of approximately \SI{32}{\micro\meter}. The total number of turns in the coil is 4360. To prevent electrical shorts between turns, each winding layer, as well as the solenoid body, was coated with a thin layer of GE Varnish, a cryogenic adhesive that provides both electrical insulation and improved thermal contact. A photo of the assembled solenoid magnet, with the CMS removed, is shown in \Cref{fig:3_3d_overview}(c).

%To evaluate the accuracy of the simulation, the magnetic field generated by the finished solenoid withing the shielding assembly was measured along the central z-axis at room temperature with the 'AS-UAP GEO-X' axial probe from Projekt Elektronik. Due to the limited space and access inside the closed shielding assembly, it was not possible to measure the field distribution in radial direction.
%The measurement results are shown in \aref{fig:2_surface_plot}\,(e).
%While the simulation of the assembly takes into account the asymmetric winding pattern, as discussed in \aref{subsection:solenoid_construction}, and predicts the shift of the maximum magnetic field strength towards the bottom of the coil, its results are not accurate enough to predict the position with the most homogeneous distribution of the magnetic field, which we attribute to the finite size of the mesh-element grid. 
%Based on the measurement results, we identify a new optimal sample position with best field homogeneity, which is displaced from the centre of solenoid along central z-axis by $\SI{-40}{\milli\meter}$ and is denoted as $\mathrm{P}_0$ in \autoref{fig:2_surface_plot}\,(a).

\begin{figure}[b]
    \includegraphics[width=1.0\columnwidth]{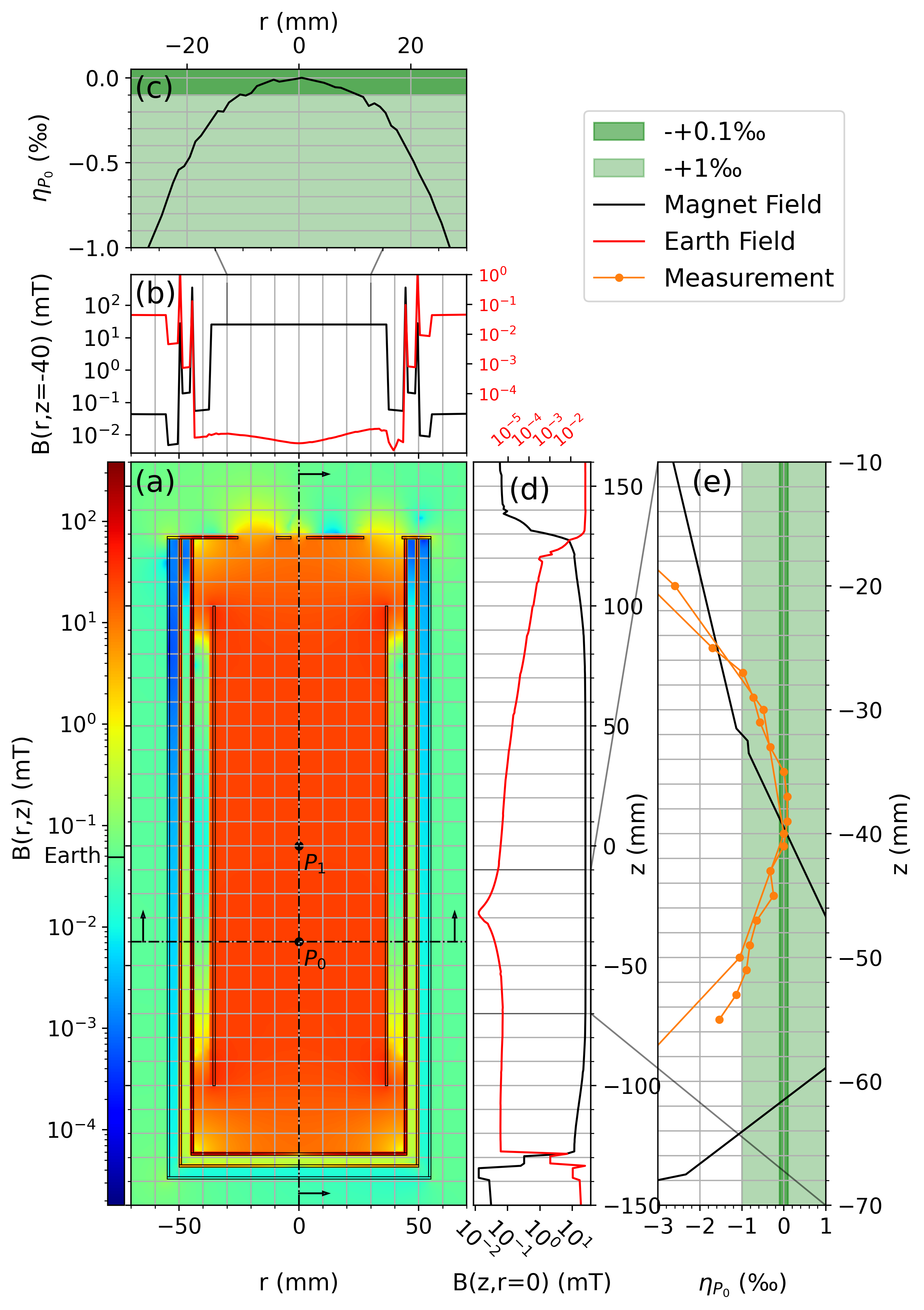}
    \caption{ 
    	(a) Vertical cross-sectional view of the assembly with the color-coded magnetic field magnitude generated by the coil for the millikelvin thermal conditions. The respective level of the Earth's magnetic field is marked on the color bar. The dots \textbf{P$_1$} and \textbf{P$_0$} mark the best sample positions according to the model calculation and the measurement, respectively. The dashed lines mark the positions along the radial direction at \textit{z}=\SI{-40}{\milli\meter} and along the vertical cylinder axis at $r=0$. The field distributions along these lines are plotted in (b) and (d), respectively.
    	The black solid lines in (b) and (d) represent the distribution of the magnetic field generated by the solenoid, and the red line shows the suppression of the ambient Earth's magnetic field.
    	In (c) and (e), the homogeneity of the magnetic field is plotted as the relative field change with respect to the magnetic field value at position \textbf{P$_0$} along radial and vertical directions, respectively.
    	In (e), the field values obtained from two room-temperature measurements are plotted as orange dots, while the black line shows the simulation result within the plotted window.}
    \label{fig:2_surface_plot}
\end{figure}

The proportionality constant between the current applied to the solenoid and the resulting magnetic field, the so-called coil constant (CC) of the magnet, has been calculated numerically from the winding geometry and then experimentally verified both at room temperature and at cryogenic conditions. A summary of the key solenoid parameters is given in \aref{tab:coil_constants}. 
A numerical calculation of the CC was used only during the initial simulation phase to determine the suitable winding pattern. 
At room temperature, the CC has been measured using an 'AS-UAP GEO-X' axial probe (same as described in \Cref{sec:field_homogeniety}) at both the geometrical center of the coil marked as position  $\mathrm{P}_1$ and the identified optimal sample position marked as $\mathrm{P}_0$ in \Cref{fig:2_surface_plot}\,(a). The CC at $\mathrm{P}_0$ is further calibrated by low-temperature electron spin resonance (ESR) measurements on a 2,2-Diphenyl-1-Picrylhydrazyl (DPPH) sample. 
DPPH is a well-known reference material used in spin resonance spectroscopy, with a widely accepted isotropic $g$-factor of 2.0037~\cite{VelluirePellat2023}. As the ESR resonance frequency $f$ depends linearly on the applied magnetic field $B$, $f = g\mu_\text{B}B/ h$ with Bohr's magneton $\mu_\text{B}$ and Planck's constant $h$, the actual magnetic field strength and, thus, CC can be precisely calibrated. We attribute the discrepancy between the CC values measured at room temperature and that determined at cryogenic temperatures to the non-linear magnetic field dependence of the permeability of the shielding material Cryophy\textsuperscript{\textregistered} as well as to a considerable error due to the small magnetic field range of only up to $\sim\SI{100}{\micro\tesla}$, which could be accessed at room temperature due to the high normal-state resistance of the superconducting coil ($\simeq\SI{4}{\kilo\ohm}$). We thus use the CC value determined from the ESR measurement on DPPH, which is the most precise value and is reproducible in subsequent measurements.

\label{sec:coil_const}
\begin{table}[tb]
   \centering
   \begin{tabular}{l|c}
	  Method  & Coil Constant (\si{\mT\per\A}) \\ %&Inductance (\si{\milli\H})\\
	   \hline
    %numerical calculation & \num{27.30}{} \\ %& \num{460}{}
    numerical simulation, $\mathrm{P}_1$   & \num{24.34(0.01)}{} \\ %& \num{456.60(0.05)}{} \\
	   measurement, RT at $\mathrm{P}_1$   & \num{23.67(0.09)}{} \\ %& \num{70(1)}{} \\
	   measurement, RT at $\mathrm{P}_0$ & \num{23.83(0.08)}{} \\ %& \num{447(1)}{} \\
	  measurement, LT at $\mathrm{P}_0$ (DPPH) & \num{24.57(0.06)}{} \\ %& \num{5.32(0.01)}{} \\

   \end{tabular}
   \caption{Values of the coil constant (CC) determined by different methods for positions P$_1$ and P$_0$.}
   \label{tab:coil_constants}
\end{table}

\section{Cryogenic Magnetic Shielding}

The cryogenic magnetic shielding (CMS) for the sample volume~2 housing the spin ensemble has been carefully engineered to fulfill two primary objectives: first, to create a magnetically quiet environment for the electronic spin ensemble, and second, to enable the generation of static magnetic fields up to \SI{50}{\milli\tesla} while minimizing their influence on the nearby sample volume~1 housing the superconducting qubits. Meeting these requirements is a key prerequisite for both the controlled operation of the spin system and keeping the magnetic field strength at a sufficiently low level to allow for optimum qubit performance. As the shielding material for the CMS, we have chosen \SI{1}{\milli\meter}-thick Cryophy$^\circledR$ sheet metal, a nickel-iron-molybdenum soft magnetic alloy known for its effective attenuation of static magnetic fields at cryogenic conditions~\cite{cryophy_datasheet}. The performance of the magnetic shields has been simulated with Comsol Multiphysics$^\circledR$. 

The whole CMS consists of three shielding cans, each being a cylinder fully closed at the bottom and open at the top. There, a copper mounting flange is attached, and a single shielding lid is added as shown by the 3D representation in \Cref{fig:3_3d_overview}\,(a). The lid has three larger holes for the magnet supports and one small hole in the center to mount the sample. The cans are separated by a gap of \SI{4}{\milli\meter} offering an inner working volume of \SI{88}{\milli\meter} in diameter and \SI{256}{\milli\meter} in height. The shielding assembly has been initially designed and fabricated with three Cryophy\textsuperscript{\textregistered} shields. However, later, the outermost Cryophy\textsuperscript{\textregistered} shield was replaced by a superconducting aluminum shield with identical dimensions, as depicted in \Cref{fig:1setup}. The replacement of the outer shield was needed for better suppression of the magnetic field level in the neighboring sample volume~1 housing the superconducting circuits, as discussed in more detail in \Cref{sec:sc_qubit_temp}. The CMS is tightly fixed to a rectangular copper flange, which is designed to match one of the side-loading RF ports of the BlueFors~XLD~1000sl system [cf. \Cref{fig:3_3d_overview}\,(a)]. 

In our Comsol$^\circledR$ simulations, the magnetic field distribution has been calculated up to a distance of \SI{1}{\meter} in radial and \SI{1.5}{\meter} in vertical direction from the magnetic shields to cover the space of the sample volume~1 hosting the superconducting qubits. The setup itself, as well as the copper elements of the sample holder, are not included in the simulation. However, we have accounted for the magnetic field dependence of the Cryophy$^\circledR$ permeability in our simulation, and a respective $B$-$H$ curve \cite{cryophy-bh-curve} has been loaded into the Comsol$^\circledR$ simulation to allow for more accurate results. We also note that the shielding performance of Cryophy$^\circledR$ is similar to that of muMetal$^\circledR$ \cite{comsol-bh-curve,mu_metal_Arpaia2021}, another popular magnetic shielding material.

\subsection{Distribution of the Magnetic Field inside the CMS}
\label{sec:field_homogeniety}

To characterize the inner shielding efficiency, we have set the ambient magnetic field outside of the shielded volume to \SI{50}{\micro\tesla} at an angle of $25.55^\circ$ to the central vertical axis to most accurately simulate the Earth's magnetic field within the laboratory frame. The possible self-shielding effect from the superconducting solenoid~\cite{Gabrieise1988} has been neglected. The shielding efficiency is defined as a ratio of the total field amplitude in the center of the solenoid, $B_{\mathrm{in}}\,(r=0,z=0)$, to the static background Earth's field, $B_{\mathrm{out}}=B_{\mathrm{Earth}}\simeq\SI{50}{\micro\tesla}$. It has been calculated for combinations of one, two, and three cans, including the shielding lid for the permeability values corresponding to the experimental cryogenic transitions. Field suppression factors of $B_{\mathrm{in}}\,(r=0,z=0)/B_{\mathrm{out}} \simeq 5\cdot10^{-3}$, $5\cdot10^{-4}$ and $1\cdot10^{-4}$ have been obtained, respectively. The residual radial and axial field strengths, $B_{\mathrm{in}}\,(r)$ and  $B_{\mathrm{in}}\,(z)$, calculated for the complete CMS assembly are plotted in \Cref{fig:2_surface_plot} (b) and (d) as the red curves, respectively, with radial coordinate $r$ and axial coordinate $z$ both having their origin in the geometrical center of the solenoid. At room temperature, the attenuated background field within the shields was measured to be \SI{0.052(6)}{\uT}, which corresponds to a reduction of the measured ambient magnetic field outside of the shields of $\simeq$\SI{48.7}{\uT} by a factor of about $10^3$. %The details on this measurement are provided in (\aref{sec:field_homogeniety}).

Taking the dimensions specified above for the solenoid and CMS, the calculated spatial distribution of the magnetic field generated by the solenoid is shown in \Cref{fig:2_surface_plot}. 
%The coil is simulated to generate a magnetic field up to \SI{50}{\milli\tesla} in its center.
The optimum position for a spin ensemble sample with the most homogeneous field distribution is marked as \textbf{P$_0$} in \Cref{fig:2_surface_plot}\,(a). It is shifted from the geometrical center of the solenoid, \textbf{P$_1$}, by \SI{-40}{\milli\meter} along the vertical $z$-axis. The field homogeneity $\eta_{P_0}(r,z)$ in the vicinity of $P_0$ is defined as
\begin{equation}
    \eta_{\mathrm{P_0}}\,(r,z) = \frac{B(r,z)}{B_{\mathrm{P_0}}}-1
\end{equation}
and plotted in \Cref{fig:2_surface_plot}\,(c) and (e). Here, $B(r,z)$ is the magnetic flux density at the position with coordinate $r$ along the radial direction and $z$ along the vertical axis, and $B_{\mathrm{P_0}}=B(r=0,~$z=\SI{-40}{\milli\meter}$)$ is the flux density at the position \textbf{P$_0$}, which represents the measured magnetic center of the coil at coordinates $r=0$ and $z=\SI{-40}{\milli\meter}$. %The details on the field distribution measurement are discussed in \Cref{sec:coil_const}. 
The simulated radial field homogeneity is shown in \Cref{fig:2_surface_plot}\,(c) and is expected to be better than $10^{-4}$ within a range of $\pm\SI{1}{\centi\meter}$ in radial direction. 

We also measured the field homogeneity along the central $z$-axis using an 'AS-UAP GEO-X' axial probe from Projekt Elektronik. These measurements were performed at room temperature for the fully assembled CMS and solenoid. The results are plotted in \Cref{fig:2_surface_plot}\,(e). Due to the limited space and access inside the closed shielding assembly, it was not possible to measure the field distribution also in the radial direction. The field distribution along the axial directions was measured both upwards and downwards along the $z$-axis, overlapping in the middle, see \Cref{fig:2_surface_plot}\,(e). The homogeneity along the $z$-axis differs between the measured and the modeled data. We note that while the numerical simulation of the assembly takes into account the asymmetric winding pattern and predicts a shift of the maximum magnetic field strength towards the bottom of the coil, the simulation was not accurate enough to predict the exact position with the most homogeneous field distribution. We attribute this to the finite size of the mesh-element grid and limitations of the calculation algorithms when including fine-size elements within a large modeled volume. Therefore, we used the measured data to reliably identify the optimal sample position with the best field homogeneity. This position is labeled $\textbf{P}_0$ in \Cref{fig:2_surface_plot}\,(a) and is displaced from the geometrical center of the solenoid (marked by $\textbf{P}_1$) along the $z$-axis by $\SI{-40}{\milli\meter}$. The magnetic field homogeneity at this position is determined to be better than $10^{-4}$ for a vertical range of $z=-40\pm\SI{5}{\milli\meter}$ and better than $10^{-3}$ for $z=-40\pm\SI{10}{\milli\meter}$ [cf. \Cref{fig:2_surface_plot}(e)]. %With this values we can reach $<10^{-4}$ magnetic field homogeneity within \SI{1}{\centi\meter^3} volume.

\subsection{Distribution of the Magnetic Field outside the CMS}
\label{sec:sc_qubit_temp}
%======================
% from Lukas's simulations files: 
% field-three_sheelds_1.055Amp_density_superconductor_outer_NK.mph - without Earth's magnetic field
% : 25mT in the magnet -- vs -- 2.4e-5 mT at the qubit postion =>  10^-6 attenuation of generated magnetic field (*10^-2 from SQQ shields)
% the superconductor is not optimally modelled....
% field-three_sheelds_1.055Amp_density_NK.mph - without Earth's magnetic field
% : 25mT in the magnet -- vs -- 2.5e-5 mT at the qubit postion =>  10^-6 attenuation of generated magnetic field (*10^-2 from SQQ shields)
% -- in the presentce of Earths field, all saturates to Earths field
%======================
Above, we have discussed how the static Earth magnetic field is reduced inside sample volume~2 by the shielding structure and derived a maximum suppression factor of $1\cdot10^{-4}$ for three concentric Cryophy$^\circledR$ shielding cans. Furthermore, the numerical simulation of the spatial distribution and attenuation of the magnetic field generated by the solenoid inside the sample volume~2 yields a reduction of the field outside the shielding structure by a factor of  $1\cdot10^{-3}$ in the presence of Earth's magnetic field  [cf. \Cref{fig:2_surface_plot}(b)] and is thus limited by the background presence of the Earth's field. Performing an additional simulation of the magnetic field distribution in the absence of the Earth's magnetic field, we find a total attenuation factor of $1\cdot10^{-6}$ at the position of the superconducting qubits in the neighboring sample volume~1 relative to the center of the coil in sample volume~2. This total attenuation factor comprises the natural spatial decay of the generated magnetic field strength (also expected to be $\sim 1\cdot10^{-3}$ from analytical model calculations \cite{Pathak.2017}) and the attenuation by the magnetic shields of sample volume~2 (expected to be $\sim1\cdot10^{-1}$ per shield). With two additional magnetic shields around sample volume~1 housing the superconducting qubits, we expect to reach a total attenuation factor of $1\cdot10^{-8}$. Then, at a field level of \SI{50}{\milli\tesla} in the center of sample volume~2, we expect a field level of only $\pm\SI{0.5}{\nano\tesla}$ at the position of the superconducting qubits in sample volume~1 with all three non-superconducting magnetic shields.

\begin{figure*}[t]
    \includegraphics[width=\textwidth]{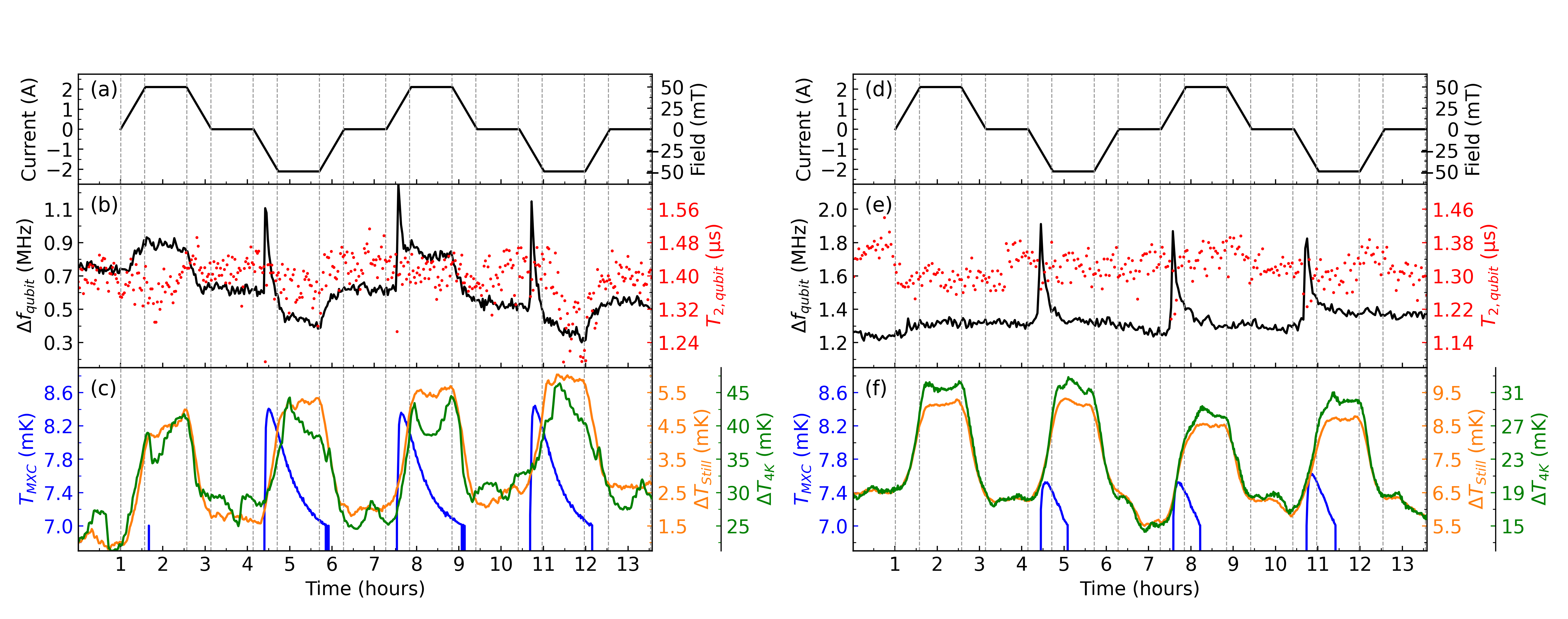}
    \caption{Time-traces of (a,d) coil current (magnetic field), (b,e) deviation $\Delta f_{\text{qubit}}$ of the flux-tunable qubit transition frequency from the value of \SI{4.877}{\GHz} at the chosen flux bias point, and (c,f) temperatures at several stages of the dilution refrigerator. The traces are measured for two experimental configurations of magnetic shields of sample volume~2: (a)-(c) with three Cryophy\textsuperscript{\textregistered} shields, and (d)-(f) with two Cryophy\textsuperscript{\textregistered} shields and one outermost superconducting aluminum shield. The temperatures for the Still and the 4\,K-stage are given as differences, $\Delta T=T-T_\text{ref}$, from reference values $T_\text{ref}$. The reference values are the following: (c) $T_{\text{ref,Still}}=\SI{1.24}{\kelvin}$ and $T_{\text{ref,4K}}=\SI{3.1}{\kelvin}$, and (f) $T_{\text{ref,Still}}=\SI{1.26}{\kelvin}$ and $T_{\text{ref,4K}}=\SI{3.1}{\kelvin}$.}
    \label{fig:4_temperatures}
\end{figure*}

To experimentally determine the actual influence of the remaining magnetic flux density in sample volume~2 on flux-tunable qubits (FTQ) in sample volume~1, simultaneous measurements of the magnetic field strength, mixing chamber temperature, qubit coherence time, and qubit transition frequency have been performed, with the result plotted in~\Cref{fig:4_temperatures}. In these measurements, two additional magnetic shields have been used in a sample volume~1 housing the FTQ, as depicted in \Cref{fig:1setup}. The flux bias point of the FTQ was chosen such that its transition frequency is most sensitive to variations in magnetic flux. We note that this operation point is associated with a short coherence time of $T_2^*\simeq$\SI{1.3}{\micro\second}. We find that there are no correlations between temporal variations of the qubit coherence time, neither with the current applied to the solenoid nor with slight variations of the base temperature of the refrigerator [cf. \Cref{fig:4_temperatures}(b) and (e)].

For the initially used configuration with three Cryophy\textsuperscript{\textregistered} shields in sample volume~2, we observe that the time-trace of the qubits transition frequency clearly follows the time-trace of the applied magnetic field, as can be seen in \Cref{fig:4_temperatures}(a) and (b). The measured change in the qubit transition frequency, $\Delta f_{\mathrm{qubit}}$,  is directly related to variations of the magnetic flux density, $\Delta B$, by the relation
\begin{equation}
	\Delta f_{\text{qubit}} = \frac{\partial f_{\text{qubit}}}{\partial \Phi} \Delta \Phi = \frac{\partial f_{\text{qubit}}}{\partial \Phi} \; S_{\text{SQUID}} \Delta B \, .
\end{equation}
Here, $\partial f_{\text{qubit}}/\partial \Phi$ is the flux sensitivity of the FTQ at the operation point and $S_{\text{SQUID}}$ the area of the SQUID loop used for flux-tuning the qubit transition frequency. We find that the $\Delta f_{\text{qubit}}$ value measured on varying the magnetic field in sample volume~2 by \SI{50}{\milli\tesla} translates into $\Delta B \simeq \SI{0.468}{\nano\tesla}$. This shows that the magnetic shielding results in a suppression of the solenoid field in sample volume~2 by a factor of $10^{8}$ at the position of the FTQ, which is in good agreement with the value derived from our simulations. 

To further improve the shielding efficiency, we have replaced the outermost Cryophy\textsuperscript{\textregistered} shield of sample volume~2 with aluminum, which is a type~I superconductor with a critical temperature of \SI{1.17}{\K}, and a critical field of approximately \SI{10}{\mT}. With this combination of two Cryophy\textsuperscript{\textregistered} and one additional superconducting aluminum shield, the field attenuation factor is further increased. This leads to a small $\Delta B$ value at the qubit position that can no longer be detected as a frequency variation of the FTQ [cf. \Cref{fig:4_temperatures}(e)]. The fast change of the magnetic flux density at the FTQ observed as noise is estimated to be as small as \SI{80}{\pico\tesla}. We note that such small magnetic flux changes may be caused by internal components of the sample volume~1, the qubits themselves, and the finite noise level of the qubit flux bias. The spikes in the FTQ frequency observed for both shielding types are due to a slight increase in the temperature of the mixing chamber plate and are observed only during the inversion of the magnetic field direction, which is not required for normal spin control. Thus, it is possible to operate the magnetic setup for spin ensembles next to FTQs without deteriorating the qubit properties.

\section{Thermal Load}
\label{sec:thermal_load_theory}

To operate the hybrid magnetic setup in a dilution refrigerator, one has to consider various sources of thermal load. They include Joule heating in the magnet's supply lines, blackbody radiation from the solenoid thermally anchored to the \SI{100}{\milli\kelvin}-stage onto the CMS attached to the MXC-stage, and the heat load during a potential quench event of the superconducting magnet.  

\subsection{Supply Lines of the Superconducting Magnet}

The supply lines of the superconducting solenoid consist of a combination of superconducting NbTi and normal conducting copper wires. The superconducting wire extends from the coil itself up to the \SI{4}{\kelvin} stage [cf. \Cref{fig:1setup}(a)]. It is thermally anchored at each intermediate stage and does not dissipate any heat. At the \SI{4}{\kelvin} stage, the superconducting wires are electrically connected to twisted pair DC lines consisting of 12 twisted pairs of \SI{35}{AWG} (\SI{0.06}{\milli\meter^2}) copper wires. Each of the two superconducting supply lines is connected to 12 copper wires. This parallel configuration allows one to safely stay within the maximum current limit of each individual copper wires.
%($I_{\mathrm{single}}\simeq0.75\cdot\SI{1750}{\mA}\simeq\SI{1310}{\mA}$) as well as for the whole bundle ($I_{\mathrm{bund}} = I_{\mathrm{single}}\sqrt{N} \simeq \SI{6400}{\mA}$ for 24 wires) when fully powering the magnet with the maximum required current of\SI{2500}{\milli\ampere}.

By shorting four twisted wire pairs at the \SI{4}{\K} stage, we have measured a total average resistance of \SI{0.57(0.02)}{\mohm} per single wire under cryogenic conditions. It is to be noted that the actual resistance of the copper wires at each thermal stage is different and gradually increases from \SI{4}{\K} to \SI{300}{\K}.
%Taking the maximal allowed current per wire of $\SI{6400}{\mA}/24\simeq \SI{270}{\mA}$, we derive a maximal dissipated power of about \SI{1}{\watt}. 
For a solenoid current of about \SI{2}{\ampere}, which is required to generate a field of \SI{50}{\milli\tesla}, we obtain an average dissipated power of less than $\SI{0.38}{\watt}$. This value is well below the cooling power of the pulse-tube refrigerator ($\simeq\SI{1.5}{\watt}$ at $\SI{4.2}{\kelvin}$ and $\simeq\SI{40}{\watt}$ at $\SI{45}{\kelvin}$). We also note that the estimated value of $\SI{0.38}{\watt}$ represents an upper limit, as it does not take into account the thermal gradient on the wire between $\SI{300}{\kelvin}$ and $\SI{4}{\kelvin}$.

By measuring the temperature at various temperature stages while ramping the magnetic field between its maximum values in opposite directions, we can extract the temperature change associated with the generated thermal load at the highest field/current values at the respective temperature stages [cf. \Cref{fig:4_temperatures}(c) and (e) for the two versions of CMS]. We observe that the temperature increase at the \SI{4}{\K}- and \SI{900}{\milli\kelvin}-stages closely follows the increase in coil current, while the temperature of the MXC-stage is not directly affected by changing the coil current. We can estimate the heat load on the \SI{4}{\kelvin}-stage due to resistive wiring of the solenoid as
\begin{equation}
    \Delta Q \simeq Q_\text{4\,K-stage} \cdot \Delta T/T_\text{4\,K-stage},
    \label{eq:cooling_pow}
\end{equation}
where $T_\text{4\,K-stage} \simeq\SI{4}{\kelvin}$ is the reference temperature of the cold stage, and $Q_\text{4\,K-stage}=\SI{1.2}{W}$ is the cooling power of the pulse tube refrigerator as provided by the manufacturer. From the measured temperature change of the \SI{4}{\K}-stage of $\Delta T \simeq \SI{15}{\milli\kelvin}$ we can estimate the heat load generated by powering the magnet to $\Delta Q \simeq \SI{4.2}{\milli\watt}$, and the operation of the solenoid does not lead to a dramatic heat load associated with the resistive wiring of the solenoid.

\subsection{Blackbody Radiation}

Due to the temperature difference between the solenoid body, which is thermally anchored to the $\SI{100}{\milli\kelvin}$-stage, and the magnetic shielding and sample assembly, which are attached to the MXC-stage with a base temperature of about $\SI{7}{\milli\kelvin}$, we need to consider the blackbody radiation from the solenoid. The power dissipated via blackbody radiation can be estimated using the Stefan-Boltzmann law:
\begin{equation}
	%0.1K^4 \cdot 5.67\cdot 10^{-8}\frac{W}{m^2\cdot K^4} \cdot 0.1m^2 = 1.45\mu W.
	P_{\mathrm{BBR}} = A \epsilon \sigma T^4,
	\label{eq:stefan_boltzmann_law}
\end{equation}
where $\sigma\simeq5.67\cdot10^{-8}\si{\watt\meter^{-2}\kelvin^{-4}}$ is the Stefan-Boltzmann constant, $\epsilon$ is the emissivity and $A$ the area of the black body. With the area $A\simeq\SI{0.1}{\m^2}$ of the solenoid surface and the emissivity $\epsilon\simeq0.07$ of copper~\cite{Kim2014,Hashmi2024}, we obtain $P_{\mathrm{BBR}}\simeq \SI{8}{\femto\W}$ at $T=\SI{100}{\milli\kelvin}$. This value is nearly three orders of magnitude smaller than the cooling power of the dilution refrigerator at the mixing chamber stage, which for BlueFors XLD~1000sl is specified to be \SI{16}{\micro\watt} at \SI{20}{\mK}. We therefore do not expect any noticeable influence of the magnet operation on the achievable base temperature of about $\SI{7}{\milli\kelvin}$.

%\subsection{Heat Load by a Quench Event}
%At the maximal applied current of \SI{2}{\A}, the energy stored in the superconducting solenoid equals \SI{0.91}{\J}. The release of this energy during a quench event will lead to a dramatic increase in the temperature of the stage, to which the solenoid is thermally anchored. To avoid potential damage to the dilution unit of the refrigerator, the superconducting solenoid is anchored to the \SI{100}{\mK}-stage instead of the MXC-stage. Thermal anchoring the magnet to stages with even higher temperatures was impossible.

\subsection{Magnetization Phenomena} 

As shown in \Cref{fig:4_temperatures}\,(c), the temperature of the mixing chamber plate, $T_{\mathrm{MXC}}$, does not follow the current/field curve. However, as shown by the blue line in \Cref{fig:4_temperatures}(c), a rapid increase of $T_{\mathrm{MXC}}$ is observed, when the polarity of current/field is inverted and its absolute values increases above $\simeq\SI{1.2}{\ampere}$ (corresponding to $\simeq \SI{21}{\mT}$). The increase in $T_{\mathrm{MXC}}$ is quite sharp and is followed by a slow, gradual decrease. We associate this observation with a remagnetization phenomenon within the Cryophy\textsuperscript{\textregistered} shields. This suggestion is supported by the reduced amplitude of the thermal spikes when the outer Cryophy\textsuperscript{\textregistered} shield is replaced by aluminum [cf. \Cref{fig:4_temperatures}(f)]. When the magnetic field (current) is increased only in positive or negative direction, temperature spikes are not observed. Similar to the estimate of the heat load on the \SI{4}{\K}-stage, we use \Cref{eq:cooling_pow} to estimate the heat load causing the thermal spikes at the mixing chamber plate. With the cooling power of $Q_{\SI{20}{\milli\kelvin}} = \SI{16}{\micro\watt}$ and an observed temperature change of $\Delta T \simeq \SI{0.8}{\milli\kelvin}$ due to the change of heat load by a removed magnetic shield, we obtain $\Delta Q_{\mathrm{1 shield}} \simeq \SI{0.72}{\micro\watt}$ and $\Delta Q_{\mathrm{3 shields}} \simeq \SI{2.16}{\micro\watt}$. This suggests that the base temperature in the absence of additional heating is at the level of \SI{6.2}{\milli\kelvin}, which is below the calibrated sensor range of \SI{7}{\milli\kelvin}.

\section{Conclusion}

In conclusion, we successfully realized a hybrid low-temperature setup that allows for the parallel operation of superconducting qubits and electron spin ensembles within a single dilution refrigerator. This has been achieved by operating the two subsystems in spatially separated sample volumes that are magnetically shielded against each other. In the sample volume hosting the spin ensemble, a stable and spatially homogeneous magnetic field in the range between \SI{-50}{\milli\tesla} and \SI{+50}{\milli\tesla} can be generated by a superconducting solenoid, which is thermally anchored to the \SI{100}{\milli\kelvin} plate of the dilution refrigerator. This field is shielded to the outside by three 1~mm-thick Cryophy\textsuperscript{\textregistered} layers, where the outermost layer is replaced by a superconducting Al shield. With the three Cryophy\textsuperscript{\textregistered} layers and the natural spatial decay of the solenoid field, the amplitude of the solenoid field at the position of the qubits is reduced by a factor of about $1\cdot 10^{-6}$. With the additional two layers of Cryoperm$^\circledR$ surrounding the qubit volume for background noise suppression, the total attenuation from the solenoid reaches a factor of $1\cdot 10^{-8}$. A further improvement has been achieved by replacing the outermost Cryophy\textsuperscript{\textregistered} layer by superconducting Al and bringing the field suppression below $1\cdot 10^{-9}$. 

The new experimental setup has been successfully tested by measuring the transition frequency and coherence time of flux-tunable superconducting qubits while varying the solenoid field up to \SI{50}{\milli\tesla}. We demonstrate that the key parameters of the qubits stay unaffected during the field sweeps at a distance of only \SI{25}{\centi\meter} between the center of the solenoid and the qubit position. Hence, our experimental setup paves the way for the study of hybrid quantum systems, integrating spin ensemble-based quantum memories with superconducting quantum processors within the same cryogenic environment, thereby enabling low-loss hybrid quantum architectures. This advancement marks a critical step toward the realization of scalable hybrid quantum systems that combine the particular strengths of spin-based memories and superconducting processors within a unified cryogenic platform.

\begin{acknowledgments}
We acknowledge support by the German Research Foundation via Germany’s Excellence Strategy (EXC-2111-390814868), and the German Federal Ministry of Education and Research via the projects QuaMToMe (Grant No.~16KISQ036) and GeQCoS (Grant No.~13N15680).
\end{acknowledgments}

%\nocite{*}
\bibliography{bibliography.bib}% Produces the bibliography via BibTeX.

\end{document}